\documentstyle[12pt,epsf]{article}
\newcommand{\stln}{\setlength{\unitlength}{2.3 ex}}

\newcommand{\sfr}{\framebox(1,1){\begin{picture}(1,1)
  \put(0,0){\line(1,1){1}}\end{picture}}}
\newcommand{\lmarcshapirowbox}
{\stln \lower2.6ex\hbox{
\begin{picture}(11.9,2.6)
\multiput(.3,1.3)(1,0){2}{\sfr}
\put(2.3,1.3){\framebox(3,1){$\cdots$}}
\put(5.3,1.3){\sfr}
\put(6.3,1.3){\sfr}
\put(7.3,1.3){\framebox(3,1){$\cdots$}}
\put(10.3,1.3){\sfr}
\put(6.3,1){$\underbrace{~~~~~~~~~~~~~~~}_{2j_{\phantom{}_L}}$}
\multiput(.3,.3)(1,0){2}{\sfr}
\put(2.3,.3){\framebox(3,1){$\cdots$}}
\put(5.3,.3){\sfr}
\put(.3,0){$\underbrace{~~~~~~~~~~~~~~~~~~}_{n}$}
\end{picture}}}
\newcommand{\rmarcshapirowbox}
{\stln \lower2.6ex\hbox{
\begin{picture}(11,2.6)
\multiput(.3,1.3)(1,0){1}{\sfr}
\put(1.3,1.3){\framebox(3,1){$\cdots$}}
\put(4.3,1.3){\sfr}
\put(5.3,1.3){\sfr}
\put(6.3,1.3){\framebox(3,1){$\cdots$}}
\put(9.3,1.3){\sfr}
\put(5.3,1){$\underbrace{~~~~~~~~~~~~~~~}_{2j_{\phantom{}_R}}$}
\multiput(.3,.3)(1,0){1}{\sfr}
\put(1.3,.3){\framebox(3,1){$\cdots$}}
\put(4.3,.3){\sfr}
\put(.3,0){$\underbrace{~~~~~~~~~~~~~~~}_{m}$}
\end{picture}}}
\let\a=\alpha

\let\l=\lambda

\let\s=\sigma

\newcommand{\be}{\begin{equation}}
\newcommand{\ee}{\end{equation}}
\newcommand{\bea}{\begin{eqnarray}}
\newcommand{\eea}{\end{eqnarray}}

\renewcommand{\paragraph}[1]{
\vspace{.8mm}\par\noindent {\sl #1}\\ \vspace{0.2mm} }

\newcommand{\ba}{\left(\begin{array}}
\newcommand{\ea}{\end{array}\right)}

\def\da{{\dot\alpha}}
\def\ddt{{\partial_\tau}}

\def\dmin{{\partial_-}}
\def\dpl{{\partial_+}}

\def\su{{SU(2,2|4)}}

\newsavebox{\uuunit}
\sbox{\uuunit} {\setlength{\unitlength}{0.825em}
 \begin{picture}(0.6,0.7)
\thinlines \put(0,0){\line(1,0){0.5}} \put(0.15,0){\line(0,1){0.7}}
 \put(0.35,0){\line(0,1){0.8}}
\multiput(0.3,0.8)(-0.04,-0.02){12}{\rule{0.5pt}{0.5pt}}
\end {picture}}

\begin{document}
\begin{titlepage}
\begin{flushright}
SU-ITP-99/22\\ KUL-TF-99/16\\PSU-TH-208\\
 {\tt hep-th/9905112}\\ May 17, 1999\\
\end{flushright}
\vskip 2cm
\begin{center}
{\LARGE {\bf Supertwistors as Quarks of $\su$}}
\vskip 1 cm
{\bf Piet Claus$^{\dagger a}$, Murat Gunaydin$^{* b}$, Renata
Kallosh$^{**c}$,\\ J.  Rahmfeld$^{**d}$ and Yonatan Zunger$^{**e}$}\\
\vskip.5cm
{\small $^\dagger$ Instituut voor theoretische fysica, \\ Katholieke
Universiteit Leuven, B-3001 Leuven, Belgium\\
$^*$ Physics Department,\\ Penn State University,
University Park, PA, 1682, USA\\
$^{**}$ Physics Department,\\ Stanford University, Stanford,
CA 94305-4060, USA\\ }
\vskip 2cm
\begin{abstract}
The GS superstring on $AdS_5\times S^5$ has a nonlinearly realized,
spontaneously broken $\su$ symmetry.  Here we introduce a two-dimensional
model in which the unbroken $\su$ symmetry is linearly realized.  The basic
variables are supertwistors, which transform in the fundamental
representation of this supergroup.

The quantization of this supertwistor model leads to the complete oscillator
construction of the unitary irreducible representations of the centrally
extended $\su$.  They include the states of $d=4$ SYM theory, massless and
KK states of $AdS_5$ supergravity, and the descendants on $AdS_5$ of the
standard massive string states, which form intermediate and long massive
supermultiplets.  We present examples of long massive supermultiplets and
discuss possible states of solitonic and $(p,q)$ strings.
\end{abstract}
\end{center}
\vfill
\footnoterule \noindent
\begin{tabular}{ll}
{\footnotesize $\phantom{a}^a$ e-mail: piet.claus@fys.kuleuven.ac.be. } &
{\footnotesize $\phantom{b}^b$ e-mail: murat@phys.psu.edu. }\\
{\footnotesize $\phantom{c}^c$ e-mail: kallosh@physics.stanford.edu. } &
{\footnotesize $\phantom{d}^d$ e-mail: rahmfeld@leland.stanford.edu. }\\
{\footnotesize $\phantom{e}^e$ e-mail: zunger@leland.stanford.edu. }&
\end{tabular}
\end{titlepage}

\section{Introduction}
Supertwistors have not yet been fully incorporated into the study of the
$AdS$/CFT correspondence \cite{Malda}.  Just as Penrose twistors
\cite{Penrose} carry the fundamental representation of the conformal group,
Ferber's supertwistors \cite{Ferber} carry the fundamental representation
of superconformal symmetry.  In this paper we shall restrict ourselves to
${\cal N}=4$ superconformal symmetry in $d=4$, which has the superalgebra
$\su$.  This is also the ${\cal N}=8$ $AdS$ algebra in $d=5$.  Supertwistor
variables, upon which $\su$ is realized linearly, may also give a natural
framework in which to study extended objects on spaces with this
symmetry.\footnote{For instance, \cite{Bandos} recently studied the
superparticle in conformal superspace with additional tensor charges using
these variables.  This paper also contains a list of important references
to twistors and supertwistors in addition to those used in the present
context.}
\par
In this note, we will present a 1+1-dimensional supertwistor model.  The
supertwistor variables will take the role of `quarks' for $\su$, generating
copies of the fundamental representation in the Fock space of states.
Using these, we derive dynamically the complete set of positive (conformal)
energy unitary irreducible representations (UIR's) of $\su$ given in
\cite{GUNMAR,Gunaydin2} from the general theory of oscillator construction
of the representations of noncompact groups \cite{GUNSAC} and supergroups
\cite{BG}.  This set includes
\begin{description}
\item[\sffamily Doubleton]  supermultiplets that  have spin range
between 2 and 4.  The
CPT--self-conjugate doubleton supermultiplet is the ${\cal N}=4$ Yang-Mills
multiplet in $d=4$ and corresponds to gauge degrees of freedom
  that couple only on the boundary of
$AdS_5$.

\item[\sffamily KK supergravity multiplets] including both the ${\cal N}=8$
graviton supermultiplet on $AdS_5$ and the massive Kaluza-Klein modes
coming from $S^5$ compactification of type IIB supergravity.  They
all have spin range 2.
\item[\sffamily General massless supermultiplets]  on $adS_5$
\footnote{  Following \cite{GUNMAR,Gunaydin2} we define massless
supermultiplets to be those obtained by taking $P=2$.}
with spin range between 2
and 4, and
\item[\sffamily `Novel' short supermultiplets] which have a non-vanishing
value of the central charge $Z$ in the centrally extended $\su$ and are
conjectured to be part of the $(p,q)$-string spectrum.  They have spin
range 2.
\end{description}
In addition to the above supermultiplets with maximum spin range 4, already
described in \cite{GUNMAR,Gunaydin2}, we will focus here on the massive
supermultiplets of $\su$ which originate from
\begin{description}
\item[\sffamily Massive string states] with maximum spin range 8.
\end{description}

The existence of this construction follows from the fact that the model has
two independent supertwistors (corresponding to left-moving and
right-moving excitations) and that independent oscillators exist at each
point along the string.  The latter condition in particular allows a Fock
space state to be built out of arbitrarily many copies of the fundamental
(`generations' in the language of \cite{BG,GUNMAR,Gunaydin2}; we will use
this notation as well).  Without this (as in the case of the particle) the
spectrum would consist only of the gauge degrees of freedom.
\par
The supertwistor action which we will present here is motivated by that of
the GS superstring on $AdS_5\times S^5$ \cite{MetsTsey,ads5}.  This theory
has a non-linearly realized, spontaneously broken $\su$ symmetry which is
inherited from the superisometries of the background space.  These
superisometries realize the algebra $\su$ nonlinearly on ten bosonic
coordinates (four $x$ along the boundary of the $AdS$ space and six $y$
perpendicular to it, or equivalently five $AdS$ directions and five
spherical angles) and sixteen fermionic coordinates $\theta$.
\par
The action of our model, on the other hand, will have a linearly realized
$\su$ symmetry which is not spontaneously broken.  The bosonic part of this
action was extracted from that of the $AdS_5\times S^5$ superstring by
restricting it to excitations only of $x$ (in effect taking
$\rho\rightarrow\infty$ and ignoring the spherical angles).  The details of
this procedure will be given in a separate paper \cite{CKRZ}, in which
various ways to realize $\su$ symmetry as isometries of supercoset spaces
will be studied.
\par
In this paper we take the point of view that a two-dimensional action with
the simplest linear realization of the $\su$ symmetry can be interesting
without necessarily being derived from any other theory.  We will focus
here on the fact that one can easily quantize this action, as there are no
gauge symmetries but only a global $\su$.

\section{Overview of the Superoscillator Construction of the UIR's of $\su$
}
The general superoscillator construction of non-compact supergroups was
presented in \cite{BG}.  The spectrum of Type IIB supergravity over the
5-sphere was first calculated in \cite{GUNMAR} and fitted into an infinite
tower of short supermultiplets of $SU(2,2|4)$.  More recently, all
doubleton and massless supermultiplets of $\su$ were given in
\cite{Gunaydin2}.
\par
The physically interesting supermultiplets are the UIR's which admit a
lowest weight state.  For these representations the spectrum of the
conformal Hamiltonian (the AdS energy) is bounded from below.
\par
The centrally extended version of the $\su$ superalgebra (which was denoted
$U(2,2|4)$ in \cite{GUNMAR}) has even subgroup $SU(2,2)\times SU(4)\times
U_Z(1)$.  The representation space is constructed using the three graded
decomposition of $\su$ with respect to its compact subsuperalgebra
$SU(2|2)_L\times SU(2|2)_R\times U(1)$
\begin{eqnarray}
[L^{0},L^{\pm}] & \subseteq & L^{\pm}\,, \cr
[L^{+},L^{-}] & \subseteq &L^{0}\,, \cr
[L^{+},L^{+}] &=& 0=[L^{-},L^{-}].
\end{eqnarray}
The generators of $SU(2,2|4)$ are given in terms of two pairs of
superoscillators
\bea
L^{-} &=& {\vec{\xi}}_{A} \cdot {\vec{\eta}}_{M}\,, \cr
L^{0} &=& {\vec{\xi}}^{A} \cdot {\vec{\xi}}_{B}
\oplus {\vec{\eta}}^{M} \cdot {\vec{\eta}}_{N}\,, \cr
L^{+} &=& {\vec{\xi}}^{A} \cdot {\vec{\eta}}^{M},\label{xieta}
\eea
where
\be
\xi_{A} = \left(\matrix{a_{i} \cr
                        \alpha_{\gamma} \cr} \right)\,,\quad
\xi^{A} = \left(\matrix{a^{i} \cr
                        \alpha^{\gamma} \cr} \right)\,
\ee
and
\be
\eta_{M} = \left(\matrix{b_{r} \cr
                        \beta_{x} \cr} \right)\,, \quad
\eta^{M} = \left(\matrix{b^{r} \cr
                        \beta^{x} \cr} \right)\,,
\ee
with $i,j=1,2$; $\gamma,\delta=1,2$; $r,s=1,2$; $x,y=1,2$ and
\be
[a_i, a^j] = \delta_{i}^{j}\,,  \quad
\{\alpha_{\gamma}, \alpha^{\delta}\} = \delta_{\gamma}^{\delta}\,,
\ee
\be
[b_r, b^s] = \delta_{r}^{s} , \quad \{\beta_{x}, \beta^{y}\} =
\delta_{x}^{y}.
\ee
Annihilation and creation operators are labeled by lower and upper
indices, respectively.  The arrows over $\xi$ and $\eta$ indicate that one
takes an arbitrary number $P$ of `generations' (copies of the  oscillators) so
that e.g.  ${\vec{\xi}}_{A} \cdot {\vec{\eta}}_{M}\equiv \sum_{K=1}^P
\xi_A(K) \eta_M(K)$.
\par
To construct a basis for a lowest weight UIR of $SU(2,2|4)$, one starts
from a set of states $ |\Omega \rangle$, referred to as the ``lowest weight
vector'' (lwv) of the corresponding UIR.  These states are defined in the
Fock space of oscillators and transform irreducibly under the maximal
compact subsupergroup $SU(2|2)_L \times SU(2|2)_R\times U(1)$.  They are
defined to be annihilated by the $L^-$ operators
\be
L^-|\Omega\rangle = 0.\label{lwv}
\ee
By acting on $ |\Omega \rangle$ repeatedly with $L^{+}$, one generates an
infinite set of states that form a UIR of $SU(2,2|4)$
\be
|\Omega \rangle ,\quad  L^{+}|\Omega \rangle ,\quad L^{+}
L^{+}|\Omega \rangle , ...
\ee
The central charge-like $U(1)_Z$ generator $Z$ which commutes with all
generators of the superalgebra $PSU(2,2|4)$, whose even subgroup is just
$SU(2,2) \times SU(4)$, is given by
\be
Z=\frac{1}{2}\{ N_{a}+N_{\alpha}-N_{b}-N_{\beta} \}.
\ee
Here $N_{a} \equiv {\vec{a}}^{i} \cdot {\vec{a}}_{i}, N_{b} \equiv
{\vec{b}}^{r} \cdot {\vec{b}}_{r}$ are the bosonic number operators,
$N_{\alpha}={\vec{\alpha}}^{\delta} \cdot {\vec{\alpha}}_{\delta} $ and
$N_{\beta} = \vec{\beta^{x}} \cdot {\vec{\beta}}_{x}$ are the fermionic
number operators.  The basis of $SU(2,2)$ given above corresponds to the
compact basis (with respect to the maximal compact subgroup $SU(2)_L\times
SU(2)_R\times U(1)$).  As was shown explicitly in the second reference of
\cite{Gunaydin2}, one can go to the noncompact $SL(2,C)\times D$ (Lorentz
group times dilatations) basis and write the generators of $SU(2,2)$ as
bilinears of four component bosonic spinors transforming covariantly under
the conformal group.  It was pointed out by these authors that the
oscillator realization can be reinterpreted in twistor language (section
2).

\section{Twistors  as quarks of the conformal group}
$SU(2,2)$ is the covering group of the conformal group $SO(4,2)$. A twistor
\cite{Penrose} is a set of two commuting, two-component spinors
which linearly realize the conformal symmetries in $3+1$ dimensions
\begin{equation}
{\cal Z}^\Omega =
\left(\begin{array}{c}\lambda_\alpha\\\mu^\da\end{array}\right)\, ; \qquad
\alpha,\,\dot\alpha =1,2\,.
\label{TwistorVector}
\end{equation}
Under $SU(2,2)$, ${\cal Z}$ transforms in the fundamental as
\be
{\cal Z}\rightarrow \pmatrix{L-\frac{i}{2} D & K \cr P & \bar
L+\frac{i}{2} D}{\cal Z} \, , \label{VectorTwistor}
\ee
where $L$ and $\bar L$ denote the two $SL(2,C)$ Lorentz transformations,
$D$ is the dilatation, $K$ is the special conformal transformation, and $P$
is the translation.
These transformations preserve the metric
\be
H=\left(\begin{array}{cc}&1\\1&\end{array}\right)
\ee
and so we can introduce a conjugate representation
\begin{equation}
\bar {\cal Z}={\cal Z}^\dagger H = (\bar\mu^\alpha , \bar\lambda_\da) \,,
\end{equation}
such that the bilinear form
\begin{equation}
\bar {\cal Z}(1) {\cal Z}(2) = \bar\mu^\alpha(1) \lambda_\alpha(2) +
\bar\lambda_\da(1)\mu^\da (2)
\end{equation}
is manifestly $SU(2,2)$--invariant.
\par
The basic idea of \cite{Penrose} is that twistors are fundamental variables
which define the space-time coordinates $x^{\dot \alpha \alpha} $ via the
relation\footnote{As usual, we replace four-vector indices with paired
spinor indices by $x^{\da\alpha}= x^\mu \bar \sigma_\mu^{\da\alpha}$.}
\begin{eqnarray}
\mu^{\da}&=&-ix^{\da\alpha}\lambda_\alpha \, . \label{mudef}
\label{fund}
\end{eqnarray}
One easily verifies \cite{Ferber} that the standard conformal
transformations of $x^{\da\a}$ follow from (\ref{VectorTwistor}).  This
suggests that twistors and supertwistors might be useful variables in the
treatment of conformal theories.
\par
It is well known that the action for a massless particle in 3+1 dimensions
can be rewritten entirely in terms of twistor variables.  Since
$P_\tau^2=0$, we can write $(P_\tau)_{\a\da}=\lambda_\a \bar \lambda_\da$,
and then using (\ref{fund}) one finds
\be
S=i\int d\tau \bar{\cal Z}\ddt{\cal Z} \,,
\label{PartActTwist}
\ee
or in components
\be
S=i \int d\tau\left(\bar\lambda_\da\ddt\mu^\da +
\bar\mu^\alpha\ddt\lambda_\alpha \right) \,.
\ee
This action is manifestly $SU(2,2)$-invariant.  Since (\ref{PartActTwist})
is quadratic it can be quantized in the usual way by imposing $[q,p]=i$:
\begin{equation}
\left[ {\cal Z}^\Omega, \bar {\cal Z}^{\Omega'}\right] =
\delta^{\Omega\Omega'}\, .
\end{equation}
In the language of \cite{GUNMAR,Gunaydin2} this gives one generation
($P=1$) of oscillators which can be used to construct the doubleton
representations of $SU(2,2)$.
\par
We would like to relate this to string theory.  For this we consider the
bosonic sector of string theory on $AdS_5\times S^5$ \cite{ads5}, and
truncate out the degrees of freedom associated with the transverse
coordinates in the limit $\rho\rightarrow \infty$, (i.e.  $\partial y=0$).
In the conformal gauge with light-cone world-sheet coordinates $
\s_\pm=\tau\pm\s$, the action in the first-order formalism becomes
\begin{equation}
S=\int d^2\sigma \left(-\frac{1}{\rho^2}(P_+)_{\alpha\da} (P_-)^{\da\alpha}
+ (P_+)_{\alpha\da}\dmin x^{\da\alpha} + (P_-)_{\alpha\da}\dpl
x^{\da\alpha}+ \dots \right)\,,
\end{equation}
with
\be
P_{\pm}^{\da\a}=\rho^2 \partial_\pm x^{\da\alpha}\,.
\label{pdef}
\ee
In the boundary limit the constraint equations become $ (P_+)^{\da\alpha}
(P_+)_{\alpha\da}= (P_-)^{\da\alpha} (P_-)_{\alpha\da}= 0 $ and can be
resolved by setting
\begin{eqnarray}
(P_+)_{\alpha\da}&=&\lambda_{L\alpha}\bar\lambda_{L\da} \nonumber\\
(P_-)_{\alpha\da}&=&\lambda_{R\alpha}\bar\lambda_{R\da} \, .
\end{eqnarray}
Using eq.  (\ref{fund}) for the left-movers and for the right-movers we can
rewrite the action as
\begin{equation}
S=\int d^2\sigma \left( i\left(\bar{\cal Z}_L\dmin{\cal Z}_L +
\bar{\cal Z}_R\dpl{\cal Z}_R\right) -\frac{1}{\rho^2}P_+P_-+...
\right)\,, \label{stringtwistor}
\end{equation}
where ${\cal Z}_{L/(R)}=(\l_{+/(-)},\mu_{+/(-)})$ as in
(\ref{VectorTwistor}), and $\mu_{+/(-)}$ is related to $\l_{+/(-)}$ as in
(\ref{mudef}).
\par
Now we see that in the limit to the boundary we have only the kinetic term
for twistors.  This free action is {\em manifestly} $SU(2,2)$ invariant.
Therefore, the theory, being a subset of the full string theory at the
boundary of the $AdS$ space, can be quantized by imposing
\begin{equation}
\left[ {\cal Z}^\Omega_\chi(\sigma,\tau), \bar {\cal
Z}^{\Omega'}_{\chi'}(\sigma',\tau')\right]_{\tau = \tau'}
=\delta^{\Omega\Omega'}\delta_{\chi\chi'} \delta(\sigma-\sigma')\, .
\end{equation}
The modes of the twistors ${\cal Z}$  are nothing but the infinite
set of oscillators used in \cite{BG,GUNMAR,Gunaydin2} to construct
representations of $SU(2,2)$. Therefore, we have found states in
string theory on $AdS_5$ which form representations of $SU(2,2)$.

\section{Supersymmetric model}
Supertwistors \cite{Ferber} realize the full $SU(2,2|4)$ algebra linearly.
They contain two anticommuting 2-component spinors $(\xi, \varepsilon)$ in
addition to the commuting spinors $(\lambda,\mu)$.  The supergroup action
on the fundamental representation of $\su$ is of the form:

\bigskip
\centerline{\epsffile{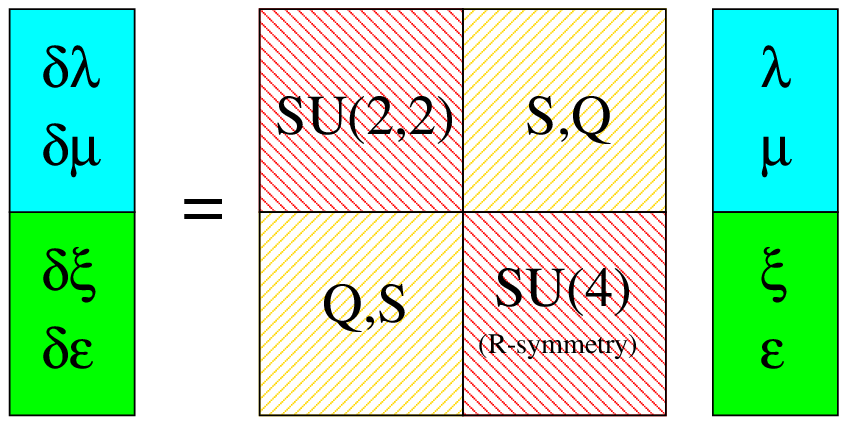}}
\medskip

Here again one considers supertwistors\footnote{Detailed notation
for supertwistors will be given in \cite{CKRZ}.}
as fundamental, and the $(x,
\theta)$-superspace as a derived concept.  The supertwistor ${\cal Z}$ and
the metric now naturally extend to
\be
{\cal Z} = \pmatrix{\lambda_\a\cr \mu^\da \cr \xi^i \cr \varepsilon^r} \,,
\qquad
H = \left(\begin{array}{cc|cc}&1&\\1&&\\\hline&&-1\\&&&
-1\end{array}\right)\,,
\qquad
\bar {\cal Z} = {\cal Z}^\dagger H \,.
\ee
We replace (\ref{fund}) by
\begin{eqnarray}
\mu^{\da}&=&-i \left (x^{\da\alpha} +  {i\over 2}  (\bar \theta^{\dot
\alpha}_i \theta^{\alpha i}  + \bar \theta^{\dot \alpha}_r
\theta^{\alpha r}) \right ) \lambda_\alpha \equiv -i z^{\da\alpha}
\lambda_\alpha\, \label{Smudef}
\end{eqnarray}
and similarly define $\theta$ by
\be
\xi^i=\theta^{\a i} \lambda_\a\, , \quad i=1,2; \qquad  \varepsilon^r =
\theta^{\a r} \lambda_\a  \quad r=1,2,
\label{split}
\ee
so that the bilinear form $\bar {\cal Z}(1) {\cal Z}(2)$
is manifestly invariant under $\su$. An additional constraint comes
from the reality of $x^\mu$, which implies that the $\su$-invariant
norm of ${\cal Z}$ vanishes.
\par
The two-dimensional supertwistor model we would like to consider is
\begin{equation}
S=i \int d^2\sigma \left(\bar{\cal Z}_L\dmin{\cal Z}_L + \bar{\cal
Z}_R\dpl{\cal Z}_R\right) \,.
\label{toy}
\end{equation}
This world-sheet action is manifestly $\su$-invariant.  The canonical
quantization condition is
\begin{equation}
\left[{\cal Z}^\Omega_\chi(\sigma,\tau), \bar {\cal
Z}^{\Omega'}_{\chi'}(\sigma',\tau')\right\}_{\tau = \tau'} =
\delta^{\Omega\Omega'}\delta_{\chi\chi'} \delta(\sigma-\sigma')\, .
\end{equation}
On-shell for left- and right-movers we may use an expansion
\begin{eqnarray}
{\cal Z}_{L}= \sum_{n=-\infty}^{ n=\infty} e^{-in(\tau + \sigma)} Z_{Ln} \ ,
\qquad {\cal Z}_{R}= \sum_{n=-\infty}^{ n=\infty} e^{-in(\tau - \sigma)}
Z_{Rn}\,.
\end{eqnarray}
The quantization condition for the left and right movers takes the form
\begin{equation}
\left[ Z^\Omega_{Lm} , \bar Z^{\Omega'}_{Ln}\right\}
=\delta^{\Omega\Omega'}\delta_{mn} \, , \qquad  \left[
Z^\Omega_{Rm} , \bar Z^{\Omega'}_{Rn}\right\}
=\delta^{\Omega\Omega'}\delta_{mn} \, ,
\end{equation}
and the Hamiltonian reads
\begin{equation}
{\cal H} = i \int d \sigma \left(\bar {\cal Z}_L {\cal Z}_L' - \bar {\cal Z}_R
{\cal Z}_R' \right) =
\sum_{n=-\infty}^{\infty}
n( \bar Z_{Ln} Z_{Ln} + \bar Z_{Rn} Z_{Rn})\,.
\end{equation}
\par
Thus we find that the world-sheet supertwistor theory (\ref{toy}) has all
superoscillators which are required to construct the positive-energy
unitary supermultiplets of the $\su$ superalgebra in a Fock
space\footnote{We should note that noncompact groups admit also non-lowest
weight type representations that can not in general be realized over a Fock
space.}.  Also, the supertwistor action and the superoscillator
construction lead to the short massless and massive supermultiplets
\cite{Gunaydin2} as well as to long and intermediate supermultiplets
related to massive string states.

\section{Physical states of the supertwistor model}
To understand the connection between the two-dimensional supertwistor
action and the oscillator construction of \cite{Gunaydin2} it is useful to
split the supertwistor into components.  The Lagrangian (\ref{toy}) is
\begin{eqnarray}
{\cal L} &= & i \left ( \bar\lambda_L\dmin\mu_L + \bar\mu_L\dmin\lambda_L
- \bar \xi_L \dmin \xi_L - \bar \varepsilon_L \dmin \varepsilon_L\right.
\\ \nonumber\\
&& \left. + \bar\lambda_R\dpl\mu_R + \bar\mu_R\dpl\lambda_R
- \xi_R \dpl \xi_R - \bar \varepsilon_R   \dpl \varepsilon_R \right)
\label{sustringtwist}\,.
\end{eqnarray}
The quantization condition in components for all left-movers $\bar Z_{Lm} ,
Z_{Ln}$ is
\begin{eqnarray}
[\lambda_{\beta m} , \bar \mu^\a_n ]&=& \delta^\a_\beta
\delta_{mn}\,,\qquad [\mu^{\dot \beta}_m , \bar \lambda_{\dot \a n }] =
\delta_{\dot \a }^{\dot \beta} \delta_{mn}\,, \\ \nonumber\\ \{
\xi^i_m, \bar \xi_{j n} \}&=& \delta_j^i \delta_{mn}\,, \qquad \{
\varepsilon ^{r }_m , \bar \varepsilon_{s n}  \}= \delta_{s}^r
\delta_{mn}
\end{eqnarray}
and analogous expressions can be given for all right-movers $\bar Z_{Rm} ,
Z_{Rn}$.  To connect this construction to the superoscillators of
$SU(2,2|2+2)$ we identify the set of pairs of superoscillators ($\xi_A ,
\xi^A, \eta_M, \eta^M$) for each generation with our supertwistors as
suggested by the model (\ref{toy}):
\begin{eqnarray}
&&\xi_A = \left (\matrix{ a_i \cr a_\gamma\cr }\right )
\Longleftrightarrow\left (\matrix{
 \lambda_{\alpha } \cr
\xi^{i } \cr }\right )\,,  \qquad \xi^A = \left (\matrix{ a^i\cr
a^\gamma\cr }\right ) \Longleftrightarrow\left (\matrix{ \bar \mu^\a
\cr \bar \xi_{i } \cr }\right )\,,\\ \nonumber\\ &&\eta_M= \left
(\matrix{ b_r\cr \beta_x\cr }\right )  \Longleftrightarrow \left
(\matrix{ \mu^{\dot \beta} \cr \varepsilon ^{s } \cr }\right )\,,
 \qquad \eta^M = \left (\matrix{
b^r\cr \beta^x\cr }\right ) \Longleftrightarrow\left (\matrix{ \bar
\lambda_{\dot \a  }\cr \bar \varepsilon_{r }\cr }\right )\,.
\end{eqnarray}
\par
Our model therefore has the complete spectrum of states obtainable by the
oscillator method \cite{GUNMAR,Gunaydin2}.  All supermultiplets described
in \cite{GUNMAR,Gunaydin2} are characterized by the value of $Z$, which is
an eigenvalue of the $U(1)_Z$ generator commuting with all other generators
of $\su$.  The states with $Z=0$ form representations of the simple
superalgebra $PSU(2,2|4)$ .

For $Z=0$ the following supermultiplets of states follow from the quantized
action (\ref{toy}):
\begin{enumerate}
\item $P=1$ {\it doubleton}.  By choosing one pair of superoscillators, or
one supertwistor, one can construct the ultra-short doubleton
supermultiplets.  The CPT-self-conjugate doubleton of $\su$ corresponds to
the ${\cal N}=4$ supersymmetric Yang-Mills supermultiplet in $d=4$.  It is
the unique irreducible doubleton supermultiplet with $Z=0$.  It is known to
decouple from the Kaluza-Klein spectrum \cite{GUNMAR,krv}.  It corresponds
to a gauge supermultiplet which couples only at the boundary of the AdS
space.
\item $P=2$ {\it `massless' AdS supermultiplets}.  These include the ${\cal
N}=8$ graviton supermultiplet on $AdS_5$, which appears in the tensor
product of the two CPT-self-conjugate doubletons.  The full description of
generic supermultiplets built out of two generations of superoscillators is
given in \cite{GUNMAR,Gunaydin2}.  The general massless supermultiplets
with $Z=0$ are those given in Table 12 of the first reference of
\cite{Gunaydin2} for which $j_L=j_R$.  They have spin range between
2 and 4.  The ten-dimensional interpretation of the intermediate and
long massless supermultiplets is an interesting open problem. We see that
 we need two supertwistors for this purpose.
Obviously the ($\sigma$-independent) zero modes of the left and
right-moving twistors give a natural choice of oscillators for $P=2$
supermultiplets.
\item All $P>2$ short massive supermultiplets of spin range two, which form
an infinite tower of states with a `massless' graviton supermultiplet of
${\cal N}=8$ $AdS_5$ ($P=2$) sitting at the bottom.  These states are the
irreducible ``CPT--self-conjugate'' short supermultiplets of IIB
supergravity compactified on $S^5$ \cite{GUNMAR} (`massive' Kaluza-Klein
modes).  The lowest weight vector of these supermultiplets is the vacuum
state annihilated by all the super-oscillators ($P$ generations).
\item All $P>2$ intermediate massive supermultiplets of spin range less
than 8 and long massive supermultiplets of spin range 8 with $Z=0$.  The
supertableaux of lowest weight vectors of supermultiplets with $Z=0$ have
equal number of boxes with respect to both $SU(2|2)_L$ and $SU(2|2)_R$.
The long massive supermultiplets appear for $P\geq 4$.
\end{enumerate}
\par
For $Z\neq 0$ the following supermultiplets of states follow from the
quantized action (\ref{toy}):
\begin{enumerate}
\item The CPT non-self-conjugate doubleton supermultiplets of
\cite{Gunaydin2} with $Z=\pm {J}$.  They are constructed from one
generation of the supertwistors ($P=1$).  These are massless
representations of the ${\cal N}=4$ Poincar\'{e} superalgebra in $d=4$.
\item The intermediate and long massless supermultiplet for $P=2$ and
$Z\neq 0$ classified in \cite{Gunaydin2}.
\item The novel short supermultiplets with $P\geq 2 $ and $Z \neq 0$, which
are not CPT--self-conjugate.  They are conjectured in \cite{Gunaydin2} to
include 1/4 BPS states of Yang-Mills theory discovered in \cite{ob} and may
also belong to the spectrum of $(p,q)$ IIB strings.  These multiplets also
naturally come from quantized supertwistors of our model and here again we
have an infinite number of such states with $P\geq 2$ due to the infinite
number of modes of the 2-dimensional theory.
\item Intermediate and long massless supermultiplets with $P=2$ that are
not CPT--self-conjugate, described in \cite{Gunaydin2}.
\item Intermediate and long massive supermultiplets with $Z\neq 0$.
\end{enumerate}

\section{The descendants of the massive string states}
 A subset of the supermultiplets discussed above come directly from
 the  ten-dimensional supergravity compactified on $AdS_5\times
S^5$.  These are all massless in the  ten-dimensional sense and therefore
belong to  supermultiplets with spin range 2 and $Z=0$. The CPT
 self-conjugate
 doubleton
supermultiplet, which has spin range 1,
decouples from the spectrum as gauge modes in the bulk.

One would also like to identify  the towers of massive supermultiplets
on $AdS_5\times S^5$ that come from the massive string states. In $d=10$,
these states are not described by classical supergravity and it is this
tower of massive states which makes all the difference between field theory
and string theory in $d=10$.
\par
Obviously, since the supertwistors are the fundamentals of the $\su$ one
should be able to find such  massive states from our quantized action, or
equivalently in the framework of the superoscillator construction
\cite{GUNMAR,Gunaydin2}.   The massive string states
 are generically expected to belong to
long multiplets with spin range 8, though some of them may belong to
intermediate massive multiplets.  In fact, such massive supermultiplets can
easily be identified.  Now the regular massive states of the IIB superstring
over $AdS_5\times S^5$ are expected to fall into massive supermultiplets
with $Z=0$ and those of $(p,q)$ strings will in general have $Z\neq 0$.  A
typical example of a long massive supermultiplet is provided by choosing a
lowest weight vector of $\su$ whose supertableau with respect to
$SU(2|2)_L\times SU(2|2)_R$ is of the form:

\begin{equation}
\label{tab2}
\left|\lmarcshapirowbox ~, ~~\rmarcshapirowbox ~\right\rangle
\end{equation}
\vskip 0.5 cm
 where we take
\begin{equation}
P\geq 4; \qquad j_L,j_R \geq 2, \qquad  n,m \geq 2.
\end{equation}
\par
The lowest weight vector above leads to a supermultiplet of fields whose
$SU(2)_L\times SU(2)_R$ quantum numbers range between
\begin{equation}
(j_L-2, j_R-2)   \qquad \longrightarrow \qquad (j_L+2, j_R+2)
\end{equation}
which corresponds to a spin range of 8 and has $Z=n-m+j_L-j_R$.
 By choosing $j_L=j_R +(m-n)$ we get long massive
supermultiplets with $Z=0$ of corresponding to the massive string states of
the regular IIB superstring.  Interestingly, both the lowest spin state and the
highest spin state have the same $AdS$ energy $E=m+n+j_L+j_R+P$.  A
detailed study and classification of intermediate and long massive
supermultiplets of $\su$ will be given elsewhere \cite{Gunaydin3}.
\par
The supermultiplets of spin range 8 can be constructed starting with 4
generations of supertwistors.  This is precisely what one would have
expected: neither a superparticle action with $P=1$ nor the zero modes of
the 2-dimensional action (both left and right movers with $P=2$) would give
us massive string states.  To have $P\geq 4$ we really need the
$\sigma$-dependence in the supertwistor, as in the usual string action in
the flat superspace.

\section{Discussion}
In conclusion, our world-sheet supertwistor action may be viewed as a
quadratic quark-type action for $\su$ supergroup.  The analogy here is that
from the fundamental representations (quarks, anti-quarks) of $SU(3)$ one
can obtain all the other representations (mesons, baryons, etc.) by a
simple tensoring procedure.  Here we have two infinite towers of
generations of fundamentals of $\su$, which provide a complete set of
states of the theory which can be described in a Fock space.
\par
One should point out that the supertwistor action (\ref{toy}) is not
derived from the GS string theory on $AdS_5$.  The relation between these
two theories is more subtle, as it is the case of linear and non-linear
sigma models in general.
\par
The simple supertwistor action, as we have seen, is capable of explaining
the spectrum of all positive energy UIR's of $\su$.  Some of these
representations can be identified with Yang-Mills theory, some are states
of supergravity including KK modes, and some novel representations do not
seem to have a clear identification in any dynamical theory studied so far.
If indeed they correspond to the $(p,q)$ states of IIB superstring theory,
as conjectured in \cite{Gunaydin2}, they may not be captured by the GS
superstring action on $AdS_5\times S^5$.  Rather one may try to study the
$SL(2, {\bf Z)}$ invariant string theory \cite{Cederwall}.  The action of
Townsend and Cederwall is also known in an arbitrary background of IIB
supergravity.  Therefore we may apply the supercoset methods and gauge-fix
$\kappa$-symmetry in this string action.  The states of this theory will be
$\su$-symmetric since this will be a symmetry of the background.  But the
$SL(2, {\bf Z)}$ symmetry of this string theory suggests the presence of
(p,q) states.  It is possible therefore that we may find some
interesting relation of this $SL(2, {\bf Z)}$ symmetric string theory to
the supertwistor world-sheet model presented in this paper.
\par
Finally, we have shown here that the long supermultiplets of the $\su$
associated with the massive string supermultiplets are also present in the
spectrum of the supertwistor model.  More detailed studies of such
supermultiplets and their relation to ten-dimensional massive string states
will be important.
\par
Thus, the fact that the supertwistor world-sheet action (\ref{toy})
codifies all information on positive energy UIR's of $\su$ which can be
obtained by the superoscillator method of \cite{Gunaydin2} suggests that
this may be an interesting direction to develop.  One is even tempted to
conclude that (super) twistors on the two-dimensional world-sheet may be
fundamental whereas the (super) space-time is a derivable concept.
\par

\

\noindent
{\bf Acknowledgements.} We had stimulating discussions with I.~Bars,
M.~Bershadsky, D.~Minic, K.~Pilch, J.~Schwarz, A.~Sen, S.~Shenker, A.~Strominger,
L.~Susskind and C.~Vafa. P.C.~thanks the Physics Department of the Stanford
University for its warm hospitality and the FWO, Belgium, for the travel grant.
R.K.~is grateful for the hospitality of the Theory Group at Harvard University,
where part of this work was accomplished.
The work of P.C.~is supported by the European Commission TMR program
ERBFMRX-CT96-0045. M.G.~acknowledges support by NSF grant PHY-9802510.
R.K., J.R.~and Y.Z.~were supported in part by NSF grant PHY-9870115.

\end{document}